\documentstyle[preprint,aps]{revtex}
\input epsf
\begin{document}
\preprint{ MAD/PH/836, IFT-P 020/94, IFUSP-P 1116}
\title{{\Huge\boldmath{$\epsilon_b$}} Constraints on Self--Couplings
of Vector Bosons}
\author{O.\ J.\ P.\ \'Eboli \cite{eboli},  S.\ M.\ Lietti  \cite{sergioII}}
\address{Instituto de F\'{\i}sica, Universidade de S\~ao Paulo, \\
C.P. 20516, 01452-990 S\~ao Paulo, Brazil}
\author{M.\ C.\ Gonzalez-Garcia \cite{concha}}
\address{Physics Department, University of Wisconsin, \\
Madison, Wisconsin 53706, USA }
\author{S.\ F.\ Novaes \cite{novaes}}
\address{Instituto de F\'{\i}sica Te\'orica,
Universidade Estadual Paulista, \\
Rua Pamplona 145, 01405-900 S\~ao Paulo, Brazil.}
\date{\today}
\maketitle

\begin{abstract}
We analyze the constraints on possible anomalous contributions to
the $W^+W^-Z$ vertex coming from non-universal radiative
corrections to the $Z \rightarrow b \bar{b}$ width. We
parametrize these corrections in terms of $\epsilon_b$ and use
the LEP data to establish the allowed values for
the anomalous triple couplings. We examine all CP conserving
effective operators that exhibit $SU(2)_L\times U(1)_Y$ gauge
invariance and do not  give any tree level contribution to the
present experimental observables.  For some of these operators
our constraints are comparable with the bounds coming
from a global fit of the oblique parameters, which evidences the
increasing relevance of the precise measurement of the
$b$--quark parameters at LEP for the search of new physics.
\end{abstract}
\newpage

\section{Introduction}

The Standard Model (SM) of electroweak interactions has so far
explained extremely well  all the available experimental
measurements  \cite{sm}. From the LEP and SLC data on $Z$
physics, we can witness the agreement between theory and
experiment at the level of 1\%, which is a striking confirmation
of the $SU(2)_L\times U(1)_Y$ invariant interactions involving
fermions and gauge bosons. However, some elements of the SM,
such as the symmetry breaking mechanism and the interaction among
the gauge bosons, have not been object of direct experimental
observation yet. In particular, the structure of the triple and
quartic vector--boson couplings is completely determined by the
non--abelian gauge structure of the model, and a detailed study
of these interactions can either confirm the local gauge
invariance of the theory or indicate the existence of new physics
beyond the SM.

Presently only hadronic colliders have directly studied the
triple vertex $W^+W^-\gamma$ \cite{present}, however, their
constraints on this coupling are very loose.  One of the main
goals of LEP II at CERN will be the investigation of the reaction
$e^+e^- \rightarrow W^+W^-$, which can furnish direct bounds on
anomalous $W^+W^-\gamma$ and $W^+W^-Z$ interactions
\cite{Dieter1}. Future hadron \cite{anohad} and $e^+e^-$,
$e\gamma$ and $\gamma\gamma$ \cite{anoNLC} colliders will also
provide information on those couplings and improve significantly
our knowledge on possible anomalous gauge-boson interactions.

Other valuable indirect sources of information on anomalous
interactions are the low energy data \cite{low} and the results
of LEP \cite{DeRujula,outros,Dieter2}, which can also constrain
substantially the possible deviations of the gauge boson
self--interactions from the SM predictions through their
contribution to the electroweak radiative corrections. So far all
these works have concentrated on the {\em universal \/} effects
associated with the presence of anomalous couplings on
gauge--boson self--energies or on vertex corrections.

An interesting place to look for non--universal one loop effects
is in  the $Z\rightarrow b \bar b$ decay, since in this case the
radiative corrections are enhanced by powers of the top quark
mass due to virtual top--bottom transitions \cite{jas}.  These
non--universal contributions to the $Z b \bar{b}$ couplings have
been parametrized in a model independent way in terms of the
parameter $\epsilon_b$ \cite{Altarelli,jas}, which is defined as
\[
\epsilon_b = \frac{ g_A^b}{ g_A^\ell} - 1 \; ,
\]
where $g_A^b$ ($g_A^\ell$) is the axial coupling of the $Z$ to
the pair $b\bar{b}$ ($\ell\bar{\ell}$).

In this paper we present our analysis of the non--oblique
constraints on the $W^+W^-Z$ coupling using the parameter
$\epsilon_b$ measured at LEP. In particular, we obtain bounds on
the ``blind'' effective operators \cite{DeRujula}, {\it i.e.\/}
the ones that do not lead to any tree level effect on the
observables, exhibiting the $SU(2)_L \times U(1)_Y$ symmetry. Our
results  are comparable with the ones obtained from a global fit
of the oblique parameters what  shows the increasing relevance of
the precise  measurement of the $b$--quark parameters at LEP in
imposing bounds on new physics. This paper is organized as
follows: Section \ref{leff} presents the effective interactions
that we have analyzed, while Sec.\ \ref{conc} contains our results and
conclusions.

\section{Effective Lagrangian Parametrization} \label{leff}

As a guide line to establish the possible anomalous $W^+W^-Z$
interactions, we impose the $SU(2)_L\times U(1)_Y$ gauge
invariance as an exact symmetry above the electroweak symmetry
breaking scale $v$. This assumption is rather natural from the
aesthetic point of view and  also allows us to build effective
Lagrangians whose quantum corrections decouple in the limit of a
large cutoff scale $\Lambda$  \cite{decou}, {\em i.e.\/} the new
contributions to observable quantities  depend at most
logarithmically on the mass scale $\Lambda$.  If this gauge
symmetry is not respected by the new physics, the  induced
one-loop corrections can diverge quadratically or quartically
with the scale $\Lambda$, and the correct treatment of this
apparent unphysical situation requires a detailed knowledge of
the underlying theory \cite{Burgess}.

In principle, we can construct linear \cite{DeRujula} and
non-linear \cite{Appelquist} realizations of the gauge symmetry.
For the sake of definiteness, we choose a linear realization of
the $SU(2)_L\times U(1)_Y$ gauge invariance and of the symmetry
breaking sector in order to allow the existence of a light Higgs
boson. In practice, our results are also applicable to the
non-linear realization of the symmetry, and we also present the
bounds on anomalous interactions in this framework. Concerning
the transformation properties of the effective operators under
discrete symmetries, we will concentrate on operators that
conserve CP since  the contribution of CP violating interactions
to the decay width $Z \rightarrow b \bar{b}$ is suppressed by
powers of $m^2_b/\Lambda^2$.

With the assumptions above the basic contents of the theory are rather
general and do not depend on further details of the underlying model:
the matter sector is formed by three fermion families and the
electroweak vector bosons are the gauge bosons of the local
$SU(2)_L\times U(1)_Y$ symmetry, which is spontaneously broken by a
scalar doublet $\Phi$. The effect of new interactions at energies
below the new physics scale $\Lambda$, is parametrized in term of an
effective Lagrangian that can be written, in general, as
\begin{equation}
{\cal L}={\cal L}_{\text{SM}}
+ \sum_d \sum_i \frac{f^{(d)}_i}{\Lambda^{(d-4)}}
{\cal O}_i^{(d)} \; ,
\end{equation}
where ${\cal L}_{\text{SM}}$ is the complete renormalizable SM
Lagrangian and the effective operators ${\cal O}_i^{(d)}$ have
dimension $d$. We restricted our analysis to effective operators
that contain new contributions to the triple gauge-boson vertex
$W^+W^-Z$ and that do not give any tree level effect on the
present available experimental data (``blind directions'').
Following the notation of Ref.\ \cite{Dieter2}, the relevant
C and P conserving operators with dimension $d=6$ are \cite{Buchmuller}
\begin{equation}
\begin{array}{ll}
{\cal O}_{WWW}^{(6)}&=\mbox{\rm Tr}[\hat W_{\mu\nu}
\hat W^{\nu\rho}\hat W_\rho^{\mu}] \; , \\
{\cal O}_{B}^{(6)}& = (D_\mu \Phi)^\dagger
\hat B^{\mu\nu} (D_\nu \Phi) \; , \\
{\cal O}_{W}^{(6)}& = (D_\mu \Phi)^\dagger
\hat W^{\mu\nu} (D_\nu \Phi) \; ,
\end{array}
\end{equation}
where $\hat B_{\mu\nu} =i(g'/2) B_{\mu\nu}$, and $\hat
W_{\mu\nu}=i(g/2) \sigma^a W^a_{\mu\nu}$ with $B_{\mu\nu}$ and $
W_{\mu\nu}^a$ being the full field strengths of the $W$ and $B$ fields
and $ \sigma^a$ denoting the Pauli matrices.

The lowest dimension operator that conserves CP, but violates C and P,
has dimension 8 and is given by
\begin{equation}
{\cal O}^{(8)}_{\tilde W} = i g^2 \left\{
[(D_\mu \Phi)^\dagger\tilde{\hat W}^{\mu\nu} \Phi] [(D_\nu \Phi)^\dagger \Phi]
-
[\Phi^\dagger\tilde{\hat W}^{\mu\nu} (D_\mu \Phi)] [\Phi^\dagger (D_\nu \Phi)]
\right\} \; ,
\end{equation}
with $\tilde  W_{\mu\nu}= \frac{1}{2} \epsilon_{\mu\nu\rho\sigma}
W^{\rho\sigma}$. We would like to point out that despite being of
higher dimension in the linear realization of the
Goldstone boson sector, this operator  is of dimension 6 in the
non-linear realization of the symmetry (see below).

Therefore, the general form of the ``blind'' effective Lagrangian
that we analyze in this work is
\begin{equation}
{\cal L}_{eff} = \frac{f_W}{\Lambda^2} {\cal O}_W^{(6)}
+ \frac{f_B}{\Lambda^2}{\cal O}_B^{(6)}
+ \frac{f_{WWW}}{\Lambda^2} {\cal O}_{WWW}^{(6)} +
\frac{f_{\tilde{W}}}{\Lambda^4} {\cal O}_{\tilde{W}}^{(8)} \; ,
\end{equation}
whose contribution to the anomalous triple gauge-boson coupling
$W^+W^-Z$ can be cast into the conventional form \cite{Dieter1}
\begin{eqnarray}
{\cal L}^{WWZ}_{eff} = &  - i e \frac{c_W}{s_W}
\left [
g^Z_1 (W^+_{\mu\nu} W^{-\mu} - W^-_{\mu\nu} W^{+\mu})Z^\nu +
\kappa_{\text{Z}} W^+_\mu W^-_\nu Z^{\mu\nu} +
\frac{\lambda_Z}{m_W^2}
W_\mu^{+\nu} W_\nu^{-\rho} Z_\rho^\mu \right.  \nonumber \\
& \left. - i g^Z_5 \epsilon^{\mu\nu\rho\sigma}
(W^+_\mu \partial_\rho W^-_\nu
- W^-_\nu \partial_\rho W^+_\mu )Z_\sigma  \right ]
\label{WWZ}
\end{eqnarray}
with
\begin{equation}
\begin{array}{ll}
\Delta g_1^Z &= g_1^Z -1 =f_W \frac{\displaystyle m_Z^2}
{\displaystyle 2\Lambda^2}
\; , \\
\Delta \kappa_Z &= \kappa_Z - 1 = \left [ f_W -s_W^2(f_B+f_W) \right]
\frac{\displaystyle m_Z^2}{ \displaystyle 2\Lambda^2}
\; , \\
\lambda_{Z} &= f_{WWW} \frac{\displaystyle 3m_W^2 g^2}
{\displaystyle 2\Lambda^2} \; , \\
g^Z_{5} &= f_{\tilde W}\frac{\displaystyle m_W^4}
{\displaystyle c^2_W \Lambda^4} \; .
\label{rel:std}
\end{array}
\end{equation}
with $s_W (c_W) =\sin (\cos ) \theta_W$ and $g= e/s_W$.

Another way of parametrizing the anomalous vector bosons
couplings is through the non--linear realization of the  $SU(2)_L
\times U(1)_Y$ symmetry, which is appropriated for the study of a
strongly interacting symmetry breaking sector \cite{Appelquist}.
In this case, we should consider a chiral Lagrangian constructed
from the dimensionless unitary matrix $U$ that belongs to the
$(2,2)$ representation of the group $SU(2)_L \times SU(2)_C$.  In
the notation of Appelquist and Wu \cite{appelwu},  the  ``blind''
directions that appear in the lowest order of the chiral
expansion are described by the Lagrangians
\begin{eqnarray}
{\cal L}_2  &=& \frac{i}{2} \alpha_2 g^\prime B^{\mu\nu}
\mbox{Tr}\left \{ T [ (D_\mu U) U^\dagger, (D_\nu U) U^\dagger] \right \}
\; , \nonumber \\
{\cal L}_3  &= &\frac{i}{2} \alpha_3 g \mbox{Tr}\left \{ W^{a\mu\nu}\sigma^a
T [ (D_\mu U) U^\dagger, (D_\nu U) U^\dagger] \right \}
\; ,  \nonumber\\
{\cal L}_9  &=& \frac{i}{4} \alpha_9 g
\mbox{Tr}\left\{ T W^{a\mu\nu}\sigma^a \right \}
\mbox{Tr}\left \{ T [ (D_\mu U) U^\dagger, (D_\nu U) U^\dagger] \right \}
\; , \\
{\cal L}_{11} &=& \frac{\alpha_{11}}{2} g \epsilon^{\mu\nu\rho\lambda}
\mbox{Tr}\left \{ T (D_\mu U) U^\dagger \right \}
\mbox{Tr}\left\{ (D_\nu U) U^\dagger W^a_{\rho\lambda} \sigma^a \right\}
\; , \nonumber
\end{eqnarray}
where the custodial symmetry breaking operator $T \equiv U
\sigma_3 U^\dagger$, and the covariant derivative of $U$ is
defined as $D_\mu U  \equiv \partial_\mu U + i (g/2)
\sigma^a W^a_\mu U -  i (g^\prime/2) U \sigma_3 B_\mu$.

In the case of anomalous $W^+W^-Z$ vertex,  it turns out that the ``blind''
effective operators of the non--linear realization of the symmetry are related
to the operators of the linear one. In fact in the unitary gauge  $(U=1)$
we can make the identifications
\begin{eqnarray}
{\cal L}_2 &=& \left . \frac{8 \alpha_2}{v^2} {\cal O}_B^{(6)}
\right |_{\Phi \rightarrow \langle \Phi\rangle} \; , \nonumber \\
{\cal L}_3 &= &\left . \frac{8 \alpha_3}{v^2} {\cal O}_W^{(6)}
\right |_{\Phi \rightarrow  \langle \Phi\rangle} \; , \nonumber \\
{\cal L}_9 &=& \left .- \frac{16 \alpha_9}{v^4} \left ( \Phi^\dagger
\hat{W}^{\mu\nu} \Phi \right ) (D_\mu \Phi)^\dagger D_\nu \Phi
\right |_{\Phi \rightarrow  \langle \Phi\rangle} \; , \\
{\cal L}_{11} &=& \left . \frac{16\alpha_{11}}{g^2 v^4}~
{\cal O}^{(8)}_{\tilde{W}} \right |_{\Phi \rightarrow  \langle \Phi\rangle}
\;,  \nonumber
\end{eqnarray}
where ${\Phi \rightarrow \langle \Phi\rangle}$ stands for the
substitution of the Higgs doublet field by its expectation value.
Notice that ${\cal L}_2$, ${\cal L}_3$, and ${\cal L}_{11}$ correspond
to ``blind'' directions of the linear realization, while ${\cal L}_9$
is related to an operator of dimension 8 that does not appear in the
leading order. The contribution of the above chiral operators can be
expressed in terms of the standard parametrization (\ref{WWZ}) as
\begin{eqnarray}
\label{rel:chi}
\Delta g_1^Z &=& \alpha_3 \frac{g^2}{c_W^2} \; , \nonumber \\
\Delta \kappa_Z &=& \left [ c_W^2(\alpha_3 + \alpha_9 )-  s_W^2 \alpha_2
\right ] \frac{g^2}{c^2_W} \; ,   \\
g_5^Z &=& \alpha_{11} \frac{g^2}{c_W^2} \nonumber
\; .
\end{eqnarray}

\section{Results and Conclusions}
\label{conc}

We are now in position to evaluate the non-universal contributions to
the decay width $Z \rightarrow f_i \bar{f_i}$ due to the anomalous
interactions described above. This process takes place through the
Feynman diagram in Fig.\ \ref{diagram} whose amplitude can be
written as
\begin{equation}
\Gamma_{Zf\bar{f}}^\mu =  i\frac{e}{4 s_W c_W}
\sum_i V_{ij} V^\dagger_{ij} F(m_j) \gamma^\mu (1-\gamma^5) \; ,
\end{equation}
where $V_{ij}=\delta_{ij}$ for leptons and it is the
Cabibbo-Kobayashi-Maskawa mixing matrix for quarks. Neglecting  the
mixings $V_{tj}$ ($j=1$, $2$) and all the internal fermions
masses but $m_{\text{top}}$, it is clear that this amplitude  will be
equal for all final fermions except for the b-quark. The
universal part, $F(0)$,  affects the values of $\Delta\rho$ and
$\sin\bar\theta_W$ and it was analyzed in Ref.\ \cite{DeRujula,outros,Dieter2}.
Non-universal effects appear in the  $Z\rightarrow b\bar{b}$
width and they can be parametrized by the parameter $\epsilon_b$
\cite{Altarelli,jas} which takes the form
\[
\epsilon_b =\Delta F
\equiv  [ F(m_{\text{top}}) - F(0)] \; .
\]

The anomalous interactions give rise to new contributions to
$\epsilon_b$,  in addition to the SM ones
($\epsilon_{b}^{\text{SM}}$), that are given, in terms of the
different parametrizations of the anomalous couplings
(\ref{rel:std}) and (\ref{rel:chi}), by
\begin{eqnarray}
\label{epsiano}
\epsilon_{b}-\epsilon_{b}^{\text{SM}} &=& \Delta\kappa_Z
\Delta F_{\kappa_Z}  +
\Delta g_1^Z \Delta F_{g_1^Z}  + \lambda_Z
\Delta F_{\lambda_Z}  +
g_5^Z \Delta F_{g_5^Z} \;  \nonumber \\
&=& f_W \Delta F_{f_W} + f_B \Delta F_{f_B} + f_{WWW} \Delta F_{f_{WWW}} +
f_{\tilde W} \Delta F_{f_{\tilde W}} \\
&=& \alpha_2 \Delta F_2 + \alpha_3 \Delta F_3 + \alpha_9 \Delta F_9
+ \alpha_{11} \Delta F_{11} \; ,
\nonumber
\end{eqnarray}
where the form factors are given by
\begin{eqnarray}
\Delta F_{\kappa_Z} &=& -\frac{ g^2 c^2_W}{ 16 \pi^2}
\left\{
\frac{ z t}{ 4 w^2}
\left [ 1 + B_0(z,w,w) \right ]
+ 2 \left [ B_0(0,0,w) - B_0(0,t,w) \right ]
\right . \nonumber \\
& & + \frac{ t(t-w)}{ 2 w^2}
\left[ B_0(z,w,w)-B_0(0,t,w) \right ]
- 2 w \left [ C_0(0,z,0,t,w,w) -  C_0(0,z,0,0,w,w) \right ] \nonumber \\
& &\left .
+t \left [ \frac{ (t-w)^2+t z- 2zw + 4w^2}{ 2 w^2}
\right ] C_0(0,z,0,t,w,w)
\right\} \; ,
\end{eqnarray}
\begin{eqnarray}
\Delta F_{g^Z_1} &=&
- \frac{ g^2 c^2_W}{ 16 \pi^2}
\left \{
-\frac{ t} { 2 w}
+\frac{ 3 t} { 2 w} B_0(z,w,w)
+2 \left ( 1+\frac{ w}{ z} \right )
\left[ B_0(0,0,w)- B_0(0,t,w) \right]
\right . \nonumber  \\
& &+2 \frac{ t}{ w}
\left ( 1+\frac{ t+ w}{ 2 z} \right )
\left[ B_0(0,t,w)- B_0(z,w,w) \right]
+2 w \left (1+\frac{ w}{ z}\right ) C_0(0,z,0,0,w,w)
\\
& & \left .
- \left [ 2 \frac{ (t-w)^2+t z +w z } { z}
+ \frac{ t} {  w}
\frac{ (t-w)^2+tz-2wz +z^2}{ z}
\right ] C_0(0,z,0,t,w,w)
\right\} \; , \nonumber \\
 & & \nonumber \\
\Delta F_{\lambda_Z} &= &
- \frac{\displaystyle  g^2 c_W^2}{ \displaystyle 16 \pi^2 w}
\left \{ 2w~ \left[ B_0(0,0,w)- B_0(0,t,w) \right ]
+ (2 w-z)t~ C_0(0,z,0,t,w,w)
\right. \nonumber \\
&  & \left.
-2w^2~ \left [ C_0(0,z,0,t,w,w)-C_0(0,z,0,0,w,w) \right ]
\right \} \; ,
\end{eqnarray}
\begin{eqnarray}
\Delta F_{g^Z_5}  &=&
\frac{ g^2 c^2_W}{ 8 \pi^2}
\left\{
\left ( \frac{ t}{ w}
         -\frac{ t}{ z} \right )~
\left[ B_0(z,w,w)- B_0(0,t,w)\right]
- \left ( 1 - \frac{ w}{ z} \right )~
\left[ B_0(0,0,w)- B_0(0,t,w)\right]
\right.  \nonumber  \\
& &  + \left[ \frac{ t(t-w)}{ w}
-\frac{ (t-w)^2}{ z} \right ]~ C_0(0,z,0,t,w,w)
+\frac{ w^2}{ z}~ C_0(0,z,0,0,w,w) \\
& & \left.
+ w~ \left[ C_0(0,z,0,t,w,w)-C_0(0,z,0,0,w,w)\right]
+\frac{ t z }{ 2 w} C_0(0,z,0,t,w,w)
\right \} \; . \nonumber
\end{eqnarray}
$B_0$ and $C_0$ are the Passarino--Veltman two and
three-point functions respectively \cite{pas:vel}. We used the
short hand notation $z=m_Z^2$, $w=m_W^2$, and
$t=m_{\text{top}}^2$ and our sign conventions are the same as
those in Ref.\ \cite{Dieter2}. The Passarino--Veltman two-point
functions are divergent and were evaluated  using dimensional
regularization that is a gauge--invariant regularization scheme. The
above expressions have poles only in $d=4$ dimensions that are
identified with the logarithmic dependence on the cutoff,
following the prescription given in Ref.\ \cite{Dieter2}, {\it
i.e.},
\begin{equation}
\frac{2}{4-d} -\gamma_E + \ln(4\pi) +1 =\ln\frac{\Lambda^2}{\mu^2} \; .
\end{equation}
The form factors $\Delta F_{\lambda_Z}$ and $\Delta F_{g^Z_5}$
are independent of the cutoff while the others have a logarithmic
dependence of the form
\begin{eqnarray*}
\Delta F_{\kappa_Z} &=&- \frac{ g^2 c_W^2}{ 64 \pi^2}
\frac{ z t}{  w^2}
\ln\frac{ \Lambda^2}{ w} \; ,\\
\Delta F_{g_1^Z} &=& - \frac{ 3 g^2 c_W^2}{ 32 \pi^2}
\frac{ t} {  w}
\ln\frac{ \Lambda^2}{ w} \; .
\end{eqnarray*}

{}From Eqs.\ (\ref{rel:std}), (\ref{rel:chi}), and (\ref{epsiano})
we obtain the form factors for the different parametrizations of
the anomalous vertex,
\begin{equation}
\begin{array}{ll}
\Delta F_{f_W}&=\left(c_W^2 \Delta F_{\kappa_Z} +\Delta F_{g_1^Z}\right)
\frac{ \displaystyle m_Z^2}{\displaystyle 2\Lambda^2} \; ,\\[+0.2cm]
\Delta F_{f_B}&= -s_W^2
\frac{ \displaystyle m_Z^2}{ \displaystyle 2\Lambda^2} \Delta F_{\kappa_Z}
\; , \\[+0.2cm]
\Delta F_{f_{WWW}} &=
\frac{ \displaystyle 3m_W^2 g^2}{\displaystyle 2\Lambda^2}
\Delta F_{\lambda_Z}  \; , \\[+0.2cm]
\Delta F_{f_{\tilde W}} &=
\frac{  \displaystyle m_W^4}{\displaystyle  c^2_W\Lambda^4} \Delta F_{g^Z_5}
\; ,
\label{Dieter1}
\end{array}
\end{equation}
and,
\begin{eqnarray}
\label{Dieter2}
\Delta F_2 &=& - g^2 \frac{ s_W^2}{ c_W^2}
\Delta F_{\kappa_Z} \; , \nonumber \\
\Delta F_3 &= &g^2 \left ( \Delta F_{\kappa_Z} +
\frac{ 1}{ c_W^2} \Delta F_{g_1^Z} \right ) \; , \nonumber  \\
\Delta F_9 &= & g^2 \Delta F_{\kappa_Z} \; , \\
\Delta F_{11} &=& \frac{ g^2}{ c_W^2}
\Delta F_{g^Z_5} \; . \nonumber
\end{eqnarray}

In order to obtain the bounds on the anomalous triple
gauge--boson vertices, we evaluated these form factors as a
function of $m_{\text{top}}$, neglecting all other fermion masses. Our
results are shown in Table \ref{formfac} which  also contains the
SM prediction for $\epsilon_b$ quoted in Ref.\ \cite{Altarelli}.
It is important to notice that the SM prediction
($\epsilon_b^{\text{SM}}$) is practically independent of the
Higgs boson mass and it was obtained assuming that
$\alpha_s(m_Z)=0.118$.

An analysis of the available LEP and SLC data in terms of the
oblique parameters $\epsilon_i$ ($i=1$, $2$, $3$) and
$\epsilon_b$ has been performed in Ref.\ \cite{altarellinew},
which led to
\begin{equation}
\epsilon_b= \left( 0.9  \pm 4.2 \right ) \times 10^{-3} \; .
\label{fit}
\end{equation}
This experimental result together with Eqs.\ (\ref{epsiano}),
(\ref{Dieter1}), and (\ref{Dieter2}) and Table \ref{formfac} can
be used to constrain the ``blind'' directions of the low--energy
effective Lagrangian in the different parametrizations. In Table
\ref{const1} we present, for several  values of the top quark
mass, the central values and 1$\sigma$ errors of the coefficients
of the four anomalous operators with a linear realization
of the $SU(2)_L \times U(1)_Y$. These results were obtained
assuming that only one coefficient is different from zero
at a time and the scale $\Lambda= 1$ TeV. Table \ref{const1nl}
contains the central values and 1$\sigma$ errors for the
coefficients of the non--linear effective operators.  For
completeness, we also list in Table \ref{const2} the bounds at
95\% CL for the parameters $g_1^Z$, $\kappa_Z$, $\lambda_Z$, and
$g_5^Z$. As expected the constraints  get  more restrictive for
large top quark masses since the form factors become larger.

It is interesting to notice that the central value of $\epsilon_b$ (\ref{fit})
is positive while the SM contribution is negative (see Table \ref{formfac}).
These two facts favour the new physics beyond the SM that  gives rise to
positive contributions to $\epsilon_b$. In our case, this leads to rather
asymmetric  constraints on the anomalous triple gauge-boson vertices --- that
is, anomalous couplings that tend to shift the value of $\epsilon_b$  in the
negative direction are strongly constrained while positive shifts are favoured.
For all couplings, except the one associated to $f_B$ (or $\alpha_2$),
this implies that negative values of the couplings are favoured as can be seen
in Tables \ref{const1}, \ref{const1nl}, and \ref{const2}.

At this point it is worth comparing our results with the bounds
obtained from the universal radiative corrections, which can be
found, for instance, in Table I of Ref.\ \cite{Dieter2}. A crude
comparison shows  that our constraints  are comparable to the
universal bounds for $f_W$. Moreover, our results for  $f_{WWW}$,
or equivalently for $\lambda_Z$, are of the same order for a
heavy top after considering the strong correlation existing in
the universal constraints.  On the other hand, our constraints
from $\epsilon_b$ on  $f_B$ are weaker due to the smallness of
the form factor associated to the operator ${\cal O}_B^{(6)}$.

As for the bounds on the C and P violating operator  $f_{\tilde
W}$, or correspondingly $g_Z^5$, our limits are of the same order
or better than the limits springing from rare $K$ decays
\cite{He}. However, we would like to point out that the limits
derived from $\epsilon_b$ are subject to less uncertainties since
rare $K$ decays have large long--distance uncertainties, and
their  constraints should be taken cautiously.

\acknowledgments

M.C. G-G is very grateful to the Instituto de F\'{\i}sica
Te\'orica of  Universidade Estadual Paulista and Instituto de
F\'{\i}sica of  Universidade de S\~ao Paulo, where part of this
work was done, for their kind hospitality. This work was
supported by the University of Wisconsin Research Committee with
funds granted by the Wisconsin Alumni Research Foundation, by the
U.S.\ Department of Energy under contract No.\ DE-AC02-76ER00881,
by the Texas National Research Laboratory Commission under Grant
No.\ RGFY93-221,  by the National Science Foundation under
Contract INT 916182, by Conselho Nacional de Desenvolvimento
Cient\'{\i}fico e Tecnol\'ogico (CNPq/Brazil), and by
Funda\c{c}\~ao de Amparo \`a Pesquisa do Estado de S\~ao Paulo
(FAPESP/Brazil).



\begin{figure}
\epsfxsize=10cm
\begin{center}
\leavevmode \epsfbox{fig1.ps}
\end{center}
\caption{Feynman diagram contributing to $\epsilon_b$ for anomalous triple
vertices.}
\label{diagram}
\end{figure}


\begin{table}
\begin{displaymath}
\begin{array}{|c|c|c|c|c|c|}
\hline
 m_{\text{top}}   & \epsilon_{b}^{\text{SM}} \times 10^3  &
\Delta F_{\kappa_Z}\times 10^3   & \Delta  F_{g_1^Z}  \times 10^3
 & \Delta  F_{\lambda_Z}\times 10^3   & \Delta  F_{g^Z_5} \times 10^3\\
\hline
 125  &  -2.82  &  -7.43  & -27.98  &  -1.68  &  -5.63\\
 150  &  -4.88  &  -9.94  & -36.93  &  -2.10  &  -6.84\\
 175  &  -7.13  & -12.62  & -46.26  &  -2.51  &  -7.97\\
 200  &  -9.79  & -15.41  & -55.77  &  -2.90  &  -9.02\\
 225  & -12.80  & -18.29  & -65.33  &  -3.26  & -10.00\\
\hline
\end{array}
\end{displaymath}
\caption{SM model predictions for $\epsilon_b$ and anomalous form
factors as a function of $m_{\text{top}}$, assuming $\Lambda = 1$ TeV.}
\label{formfac}
\end{table}

\begin{table}
\begin{displaymath}
\begin{array} {|c|c|c|c|c|}
\hline
m_{\text{top}}   &\lambda_Z=\frac{3 m_W^2g^2}{2\Lambda^2}f_{WWW}
&  f_W \frac{ m_W^2 }{2\Lambda^2}   &  f_B \frac{ m_W^2}{2\Lambda^2}
  & g^Z_5=\frac{m_W^4}{c^2_w\Lambda^4} f_{\tilde W} \\
\hline
 125 &   -2.2\pm  2.5 &   -0.085\pm  0.096 &    1.7\pm   1.9 &
   -0.66\pm  0.75\\
 150 &   -2.8\pm  2.0 &   -0.10\pm  0.073 &    2.0\pm   1.4 &
   -0.85\pm  0.61\\
 175 &   -3.2\pm  1.7 &   -0.11\pm  0.058 &    2.1\pm   1.1 &
   -1.0\pm  0.53\\
 200 &   -3.7\pm  1.5 &   -0.12\pm  0.048 &    2.3\pm   0.92 &
   -1.2\pm  0.47\\
 225 &   -4.2\pm  1.3 &   -0.13\pm  0.041 &    2.5\pm   0.77 &
   -1.4\pm  0.42\\
\hline
\end{array}
\end{displaymath}
\caption{Low energy constraints on anomalous interactions with the
symmetry realized linearly, assuming $\Lambda = 1$ TeV.}
\label{const1}
\end{table}

\begin{table}
\begin{displaymath}
\begin{array} {|c|c|c|c|c|}
\hline
m_{\text{top}}   & \alpha_2
& \alpha_3
& \alpha_9
& \alpha_{11}
\\
\hline
 125 & 4.2 \pm   4.7 &  -0.21\pm  0.24 &   -1.3\pm  1.4 &  -1.3\pm 1.4 \\
 150 & 4.9\pm   3.5 &   -0.25\pm  0.18 &   -1.5\pm  1.1 &  -1.6 \pm 1.2\\
 175 & 5.3\pm   2.8 &   -0.28\pm  0.14 &   -1.6\pm  0.83 & -1.9\pm  1.0\\
 200 & 5.8\pm   2.3 &   -0.30\pm  0.12 &   -1.7\pm  0.68 & -2.3 \pm 0.90\\
 225 & 6.3\pm   1.9 &   -0.33\pm  0.10 &   -1.9\pm  0.58 & -2.6\pm  0.81\\
\hline
\end{array}
\end{displaymath}
\caption{Low energy constraints for the anomalous interactions with
a non-linear realization of the symmetry, assuming $\Lambda = 1$ TeV.}
\label{const1nl}
\end{table}

\begin{table}
\begin{displaymath}
\begin{array} {|c|c|c|c|c|}
\hline
   m_{\text{top}}      & \Delta\kappa_{z}   &\Delta g^Z_1   &    \lambda_Z
& g^Z_5  \\
\hline
 125 &   (-1.4,   0.43) &   (-0.38,   0.11) &   (-6.3,   1.9) &
  ( -1.9,   0.56)\\
 150 &  ( -1.3,   0.11) &   (-0.34,   0.030 )&  ( -6.0,   0.53) &
  ( -1.9,   0.16)\\
 175 & ( -1.2,  -0.091) &   (-0.32,  -0.025 )&  ( -6.0,  -0.46 )&
   (-1.9,  -0.14)\\
 200 & (  -1.1,  -0.25) &  ( -0.32,  -0.068) &  ( -6.1,  -1.3 )&
  ( -1.9,  -0.42)\\
 225 &  ( -1.1,  -0.37 )&  ( -0.32,  -0.10) &  ( -6.3,  -2.1 )&
  ( -2.1,  -0.68)\\
\hline
\end{array}
\end{displaymath}
\caption{95 \% CL limits on the anomalous couplings, assuming
$\Lambda = 1$ TeV.}
\label{const2}
\end{table}


\begin{references}

\bibitem[*]{eboli} E-mail: EBOLI@USPIF.IF.USP.BR (InterNet);
47602::EBOLI  (DecNet).

\bibitem[\dag]{sergioII} E-mail: LIETTI@USPIF.IF.USP.BR (InterNet);
47602::LIETTI  (DecNet).

\bibitem[\ddag]{concha} E-mail: CONCHA@WISCPHEN (BitNet);
47397::CONCHA  (DecNet).

\bibitem[\S]{novaes} E-mail: NOVAES@IFT.UESP.ANSP.BR
(InterNet); 47553::NOVAES (DecNet).

%
\bibitem{sm} See for instance L.\ Rolandi in Proceedings of the
{\it XXVI International Conference of High Energy Physics},
Dallas, USA, 1992
%
\bibitem{present} UA2 Collaboration, J.\ Alitti {\em et al.\/}, Phys.\ Lett.\
{\bf B277} (1992) 194; U.\ Baur and E.\ Berger, Phys.\ Rev.\ {\bf D41} (1990)
1476.
%
\bibitem{Dieter1}
K.\ Hagiwara, K.\ Hikasa,  R.\ D.\ Peccei, and D.\ Zeppenfeld
Nucl.\ Phys.\ {\bf B282} (1987) 253.
%
\bibitem{anohad}See for instance H.\ Kuijf {\em et al.\/} in {\it
Proceedings of  the ECFA Large Hadron Collider Workshop}, Acheen,
Germany,  1990 (CERN Report N0. 90-10).
%
\bibitem{anoNLC} See for instance W.\ Beenaker {\em et al.} in
{\it Proceedings of the Workshop $e^+ e^-$ Collisions at 500 GeV:
The Physics Potential} Munich-Annecy-Hamburg, 1993.
%
\bibitem{low}
A.\ Grau and J.\ A.\ Grifols, Phys.\ Lett.\ {\bf B154} (1985) 283;
J.\ C.\ Wallet, Phys.\ Rev.\ {\bf D32} (1985) 813;
W.\ J.\ Marciano and A.\ Queijeiro, Phys.\ Rev.\ {\bf D33} (1986) 3449;
J.\ A.\ Grifols, S.\ Peris, and J.\ Sol\'a, Int.\ J.\ Mod.\ Phys.\ {\bf A3}
(1988) 225;
K.\ Numata, Z.\ Phys.\ {\bf C52} (1991) 691;
S.\ Godfrey and H.\ K\"onig, Phys.\ Rev.\ {\bf D45} (1992) 3196;
F.\ Boudjema {\em et al.\/}, Phys.\ Rev.\ {\bf D43} (1991) 2223;
X-G.\ He and B.\ McKellar, Phys.\ Lett.\ {\bf B320} (1994) 165.
%
\bibitem{DeRujula}
A.\ De Rujula, M.\ B.\ Gavela, P.\ Hernandez and E.\ Masso,
Nucl.\ Phys.\ {\bf B384} (1992) 3;
%
\bibitem{outros}
A.\ Dobado, D.\ Espriu, and M.\ J.\ Herrero, Phys.\ Lett.\ {\bf B255}
(1991) 405;
P.\ Hern\'andez and F.\ J.\ Vegas, Phys.\ Lett.\ {\bf B307} (1993) 116;
G.\ B\'elanger, F.\ Boudjema, and D.\ London, Phys.\ Rev.\ Lett.\
{\bf 65} (1990) 2943.
%
\bibitem{Dieter2}
K.\ Hagiwara, S.\ Ishihara,  R.\ Szalapski and D.\ Zeppenfeld,
Phys.\ Lett.\ {\bf B283} (1992) 353;
Phys.\ Rev.\ {\bf D48} (1993) 2182.
%
\bibitem{jas} J.\ Bernabeu, A.\ Pich and A.\ Santamaria,
Phys.\ Lett.\ {\bf B200} (1988) 569; Nucl.\ Phys.\ {\bf B363} (1991) 326.
%
\bibitem{Altarelli} G.\ Altarelli, R.\ Barbieri, and F.\ Caravaglios,
Nucl.\ Phys.\ {\bf B405}  (1993) 3.
%
\bibitem{decou}
T.\ Appelquist and J.\ Carazzone, Phys.\ Rev.\ {\bf D11} (1975) 2856.
%
\bibitem{Burgess} C.\ P.\ Burgess and D.\ London, Phys.\ Rev,\ Lett.\ {69}
(1992) 3428.
%
\bibitem{Appelquist}
T.\ Appelquist and C.\ Bernard, Phys.\ Rev.\ {\bf D22} (1980) 200; A.\
Longhitano, Phys.\ Rev.\ {\bf D22} (1980) 1166 and Nucl.\ Phys.\ {\bf
B188} (1981) 118.
%
\bibitem{Buchmuller} C.\ J.\ C.\ Burges and H.\ J.\ Schnitzer, Nucl.\ Phys.\
{B288} (1983) 464;
W.\ Buchm\"uller and D.\ Wyler, Nucl.\ Phys.\ {\bf B268}
(1986) 621.
%
\bibitem{appelwu} T.\ Appelquist and G.-H.\ Wu, Phys.\ Rev.\ {\bf D48}
(1993) 3235.
%
\bibitem{pas:vel} G.\ Passarino and M.\ Veltman, Nucl.\ Phys.\
{\bf B160} (1979) 151.
%
\bibitem{altarellinew} R.\ Barbieri, talk given at Reencontres Physique
de la Valle d'Aosta, La Thuile, preprint IFUP-TH 28/94.
%
\bibitem{He} Xiao-Gang He, Phys.\ Lett.\ {\bf B319} (1993) 327.
%
\end{references}
\end{document}